%% file: hicss.tex
\documentclass[10pt]{article}

\input{hicss-packages.tex}

\usepackage{hyperref}
\usepackage{makecell}
\usepackage{graphicx}
\usepackage{comment}
\usepackage{listings}
\usepackage{todonotes}
\usepackage{comment}
\usepackage{caption}

\usepackage{algorithm}
\usepackage{algorithmic}

\title{Streamlining Attack Tree Generation: A Fragment-Based Approach}

\author{Omitted due to the double-blind review \\}

\author{Irdin Pekaric \\
University of Liechtenstein \\
{\underline{\href{mailto:irdin.pekaric@uni.li}{irdin.pekaric@uni.li}}} \\ \\
Raffaela Groner \\
Ulm University \\
{\underline{raffaela.groner@uni-ulm.de} } \\ \\
Michael Felderer\\
German Aerospace Center \\
{\underline{michael.felderer@dlr.de} } \\ \\ \And
Markus Frick \\
University of Innsbruck \\
{\underline{markus.frick@student.uibk.ac.at}} \\ \\
Thomas Witte\\
Ulm University \\
{\underline{ thomas.witte@uni-ulm.de} } \\ \\
Matthias Tichy\\
Ulm University \\
{\underline{matthias.tichy@uni-ulm.de}} \\ \\ \And
Jubril Gbolahan Adigun \\
University of Innsbruck \\
{\underline{ jubril.adigun@uibk.ac.at} } \\ \\ 
Alexander Raschke\\
Ulm University \\
{\underline{alexander.raschke@uni-ulm.de} } \\ \\ }

\date{\today}

\begin{document}

\lstdefinelanguage{AttackGraph}{
  basicstyle=\ttfamily,
  keywords={AttackTarget, AttackStep},
  keywordstyle=\color{blue}\bfseries,
  ndkeywords={OR, CPE, CWE, cweNotes, BaseScore, ImpactScore, ExploitabilityScore, epss, description, CVE, CVSS, SAND },
  ndkeywordstyle=\color{darkgray}\bfseries,
  alsoletter={CVE-, CVSS:, CWE-}
  identifierstyle=\color{black},
  sensitive=false,
  comment=[l]{//},
  morecomment=[s]{/*}{*/},
  commentstyle=\color{purple}\ttfamily,
  stringstyle=\color{red}\ttfamily,
  morestring=[b]',
  morestring=[b]",
  tabsize=2
}

\lstdefinelanguage{AttackGrammar}{
  basicstyle=\ttfamily,
  keywords={AttackTreeElement, AttackTarget, AttackStep, Model, AttackTreeModel, SubTree, Gate, AttackTreeSubElements, AttackTree, Synonyms, CVSSVECTORList, ScoreList},
  keywordstyle=\color{blue}\bfseries,
  ndkeywords={cwe, name, cpe, synonyms, cvss, note, baseScore, impactScore, exploitabilityScore, epss, attackTree, trigger, primary, numberOfDisrubtions, description, gate, probability, step, subTree, cve, attackTreeModel,cvssList, score},
  ndkeywordstyle=\color{darkgray}\bfseries,
  identifierstyle=\color{black},
  sensitive=true,
  comment=[l]{//},
  morecomment=[s]{/*}{*/},
  commentstyle=\color{purple}\ttfamily,
  stringstyle=\color{red}\ttfamily,
  morestring=[b]',
  morestring=[b]",
  tabsize=2
}

\maketitle
\begin{abstract}
Attack graphs are a tool for analyzing security vulnerabilities that capture different and prospective attacks on a system. As a threat modeling tool, it shows possible paths that an attacker can exploit to achieve a particular goal. However, due to the large number of vulnerabilities that are published on a daily basis, they have the potential to rapidly expand in size. Consequently, this necessitates a significant amount of resources to generate attack graphs. In addition, generating composited attack models for complex systems such as self-adaptive or AI is very difficult due to their nature to continuously change. In this paper, we present a novel fragment-based attack graph generation approach that utilizes information from publicly available information security databases. Furthermore, we also propose a domain-specific language for attack modeling, which we employ in the proposed attack graph generation approach. Finally, we present a demonstrator example showcasing the attack generator's capability to replicate a verified attack chain, as previously confirmed by security experts.



\end{abstract}

\subsubsection*{Keywords:}

attack trees, attack chains, DSL, attack modeling 

\section{Introduction}
\label{sec:introduction}

The process of security assurance of a system can become exceedingly complex due to various architectures and software components that need to be evaluated. In addition, systems' sizes are becoming tremendous, which makes the overall assurance and testing time-consuming and expensive. As a result, it is necessary to prioritize the crucial points or components that need to be protected and identify possible attack paths that adversaries may enforce. Otherwise, this may lead to very limited protection, allowing attackers to freely ransack the system.

With the growing complexities of cyber-physical systems (CPSs) with respect to security threats, it has become imperative to investigate and develop novel security assurance approaches~(\cite{SKANDYLAS2021421}). This includes approaches for complex systems, such as self-adaptive and AI systems that have self-management capabilities. This gives them the ability to adapt to changes in their environment or requirements without human intervention. In this regard, \cite{SKANDYLAS2021421} proposed a formal method to model the threats associated with self-protecting systems. Similarly, to capture the notion of safety concerns stemming from security attacks, \cite{witte22} postulated the design of a combined approach for modeling the safety and security aspects of self-adaptive systems. This was intended to capture the inherent vulnerabilities and risks of hazards in these systems in the form of Attack-Fault-Trees (AFTs). 

In order to provide more efficient means for security assurance, it is necessary to automate various procedures and processes. Despite some of them being very difficult to automate, for instance, in penetration testing, there are other activities that could potentially be conducted more proficiently. An example is threat modeling, or more specifically attack modeling, which provides an overview of the possible attacker actions. These models can serve as useful input for penetration testers and security experts by providing them with the needed intelligence. Furthermore, this ensures their efforts are consolidated in the right places (\cite{pekaric2023systematic}). With that said, there are still several limitations in relation to the generation and analysis of threat models. ~\cite{Shandilya2014} highlight some of the challenges associated with using attack graphs for security systems. Attack graph generation for such systems must be scalable and efficient. On the other hand, a generated attack graph must avail the analyses of and formulation of security properties and violation detection, such that the impact on the system is known and it is able to express the desired level of security assurance for the system.

\begin{figure}[htbp]
    \centering
    \includegraphics[width=\linewidth]{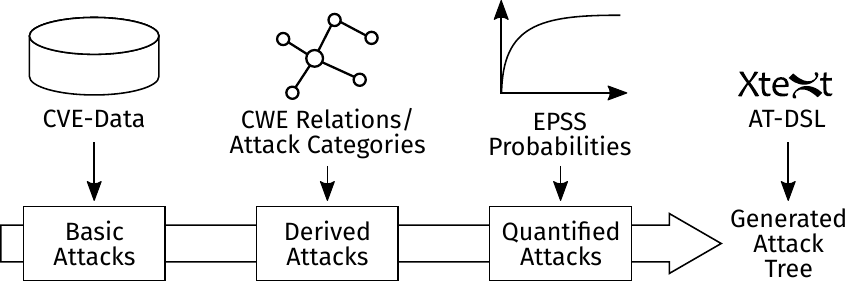}
    \caption{Overview of the attack generation.}
    \label{fig:overview}
\end{figure}

In this paper, the aforementioned research gap is addressed by proposing a novel fragment-based attack graph generation approach relying on a domain-specific language (DSL). The DSL is used to specify concepts and rules for the generated models that are in the form of attack trees. The proposed approach also provides a means of combining the generated fragments into more complex attacks by forming attack chains. An overview of the proposed generation method is presented in Figure~\ref{fig:overview}.

Our developed approach can be applied to a vast majority of CPSs including the self-adaptive and ML-enabled systems. The main idea is to generate a large number of smaller "tree" models that could be combined later by considering existing attack patterns to form more extensive models. We demonstrate our approach by providing a working example that generates an existing publicly identified attack chain. 

The remainder of this paper is organized as follows: Section  \ref{sec:background} provides background information on attack modeling and security databases. Section \ref{sec:related} outlines the related work on attack modeling and attack model generation. Section \ref{sec:approach} outlines the DSL that was developed in order to present artifacts of the generation approach for attack graphs. In addition, it explains in detail the generation approach and how the attack trees are generated. Section \ref{sec:disseval} provides a demonstrator example and a discussion of the proposed approach. Finally, Section \ref{sec:conclusion} concludes this paper. 

\section{Background}
\label{sec:background}

This section covers the background information on attack modeling techniques. The public information security databases utilized in the presented approach are also discussed.

\subsection{Attack Modeling}

\begin{figure}[]
    \centering
	\includegraphics[clip,width=\linewidth]{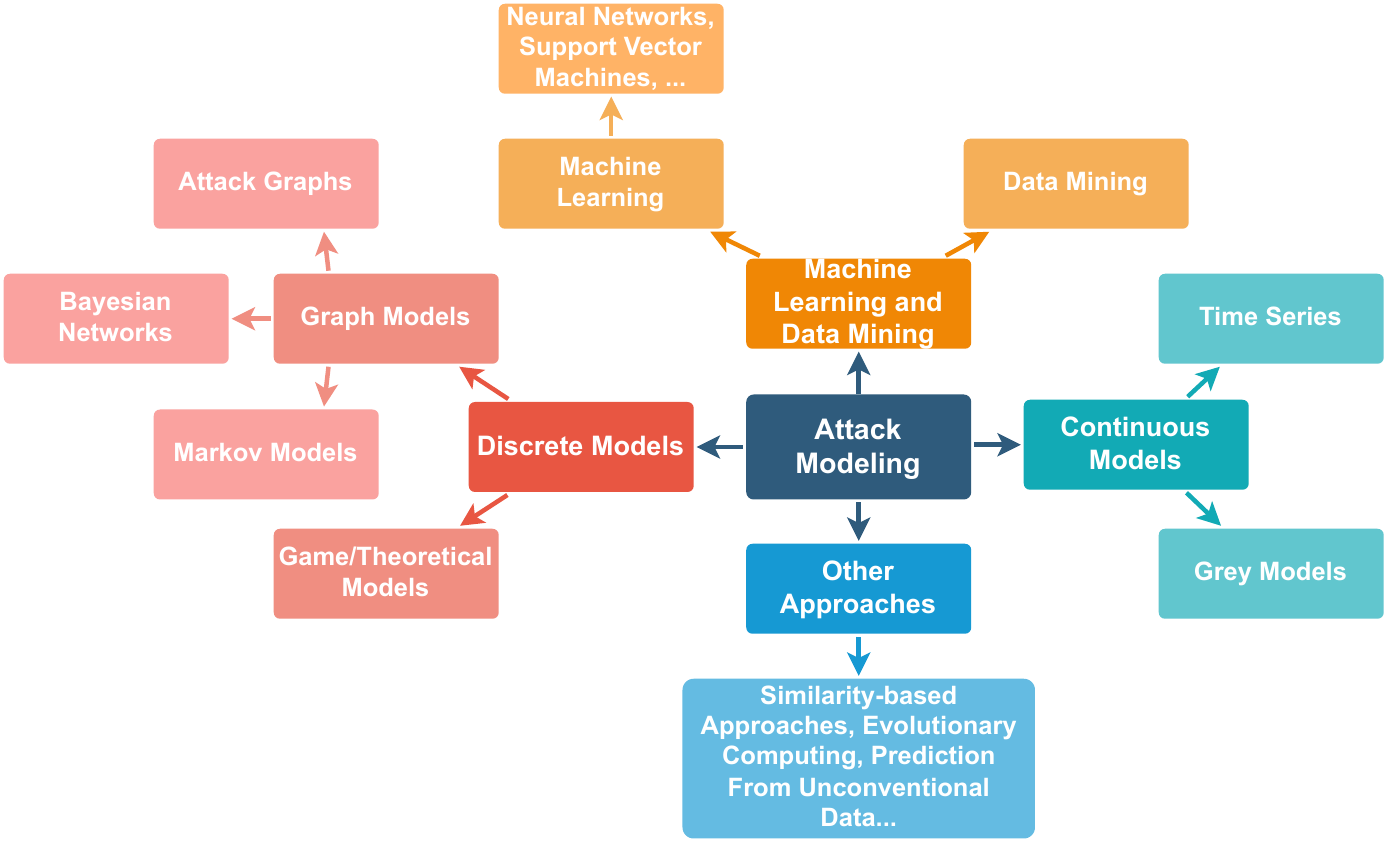}
	\caption{Attack Modeling Categories (\cite{husak2018survey})}
	\label{fig:b_attack_modeling}    
\end{figure}

Attack models portray various structures that include security-related information, such as attacks, exploits, threats, assets, vulnerabilities, and countermeasures. The  information included in the model depends on the overall goal of the model. For example, if the focus is only to have an overview of attacker actions and targets, it is not necessary to include countermeasures. In general, attack models are used to provide an overview of a system’s or a component’s security and are used to enable further actions and analysis. This allows security experts to have better insights regarding potential threats and what are the most likely targets of adversaries. \cite{husak2018survey} describe attack models as ”Prediction and Forecasting Methods in Cyber Security” and introduce their classification scheme according to which they are classified into four main categories (1) discrete models, (2) machine learning and data mining models, (3) continuous models and (4) other models. Figure \ref{fig:b_attack_modeling} presents the aforementioned classification.

In this paper, the focus is on discrete models in the form of attack graphs (\cite{haque2017evolutionary}). These can be cyclic and acyclic graphs that outline a mapping of possible real attack scenarios and attacker actions. Attack graphs are usually presented in the form of various attack paths that an attacker can execute in order for an attack to be successful. For example, this can include a variety of exploits that can target multiple vulnerabilities identified in the \texttt{ssh} library. Thus, attack graphs represent a foundation for further development and generation of attack models. This is achieved by extending attack graphs with additional information or methods, such as adding conditional variables, probabilities, states, and transitions.

More specifically, the proposed approach addresses attack trees. Attack trees consist of a root node, followed by intermediate nodes that represent the steps an attacker could take to reach the goal. Each intermediate node can have multiple child nodes, representing the various options available to the attacker at that step. In order to provide connections and relations between the nodes within the tree, a variety of logic gates can be used to form attack chains. For example, an \texttt{OR} gate can signify the choice between exploiting several vulnerabilities. Similarly, an \texttt{AND} gate denotes that each node needs to be satisfied in order for an attack to be successfully executed. Finally, the leaf nodes of the tree outline the first steps the attackers have to take to achieve their goal.

\subsection{Public Information Security Databases}

In order to uniquely distinguish between various vulnerabilities, the Common Vulnerabilities and Exposures (CVE)\footnote{\url{https://cve.mitre.org/}} data is used. Each vulnerability or CVE entry has its own unique ID, standardized scores, and description. With the aim of providing CVEs with some form of hierarchy and grouping them based on what they affect, a CWE\footnote{\url{https://cwe.mitre.org/}} representation was created. Each CWE represents a specific higher-level group or a weakness type to which all the CVEs are assigned. 

In order to provide CVE data with various metrics, the Common Vulnerability Scoring System (CVSS)\footnote{\url{https://www.first.org/cvss/}} vector was created. It provides qualitative scores related to the severity of vulnerability data. Within the vector, three metric types are found: base, temporal, and environmental. Each metric has a score that ranges from 0 to 10. In the large pool of available CVEs, it becomes very important to be able to tell which elements (system, software, platform) they affect. This is achieved using Common Platform Enumeration (CPEs)\footnote{\url{https://nvd.nist.gov/products/cpe}}, which are represented using syntax for Uniform Resource Identifiers (URI). Each CPE entry also includes the exact version of a software or library, which can be related to one or more vulnerabilities. 

CIA triad describes a standardized form of representing core information security principles (\cite{samonas2014cia}). These principles are considered substantial because they are part of the CVSS vector. That means that each CVE has scores assigned for all of the aforementioned principles. In order to conduct attacks on various types of systems, attackers exploit one or multiple vulnerabilities (CVEs). By exploiting multiple vulnerabilities to achieve a specific goal, such as obtaining root access to a system, attack chains are created. Many of the applied chains exploit CVEs that affect the same CPE. These can be identified by investigating the relations between different CWEs and propagating these links to CVEs too. Another approach to achieve the same goal is to investigate existing attack patterns, such as Common Attack Pattern Enumeration and Classification (CAPEC) \footnote{\url{https://capec.mitre.org/}}.


\section{Related Work}
\label{sec:related}

This section presents the related work on attack modeling and attack model generation. It becomes obvious that the syntax and semantics of attack graphs/trees are interpreted in different ways, thereby making it difficult to compare the different approaches. 

\cite{kotenko2013computer} present one of the core works on attack graphs. They propose an approach for attack modeling and security evaluation for
Security Information and Event Management (SIEM) systems that utilize attack graphs. These are used to model attackers' behavior, provide real-time alerts, and develop assessment procedures. This is achieved with the consideration of various security metrics, which allows the development of the Attack Modeling and Security Evaluation Component (AMSEC). This component provides additional security evaluation capabilities to the SIEMS. The overall approach consists of six steps: (1) data repository updater, (2) specification generator, (3) malefactor modeler, (4) attack graph generator, (5) security evaluator, and (6) report generator. The results include the list of possible vulnerabilities, attack trees, prediction of adversary actions, attack metrics, and suggestions on how to increase security based on the existing security policies and tools.
Whereas this work builds the attack graphs upon so-called malefactors and network topology, we use in our work CPE information gained from system analysis and crawl the CVE-DBs for corresponding vulnerabilities. Together with CWE information, our attack graphs are generated. 

\cite{al2019a2g2v} propose an approach for automated attack graph generation for computer and SCADA networks (A2G2V). It utilizes existing model-checking and architecture description tools to devise possible attack paths that can be exploited from atomic vulnerabilities. In addition, they engineer a method that parses and encodes the counterexamples as well as iterates through them until all possible attack sequences are identified. Finally, they also make it possible to visualize the generated attack graphs. Compared to our approach, their attack trees are generated on the basis of various security properties instead of publicly available security databases. These are also not generated based on a DSL and cannot be combined with other trees. 

\cite{li2022attackg} introduce an approach for constructing attack graphs from cyber threat intelligence (CTI) reports. This is achieved with the help of structured threat intelligence defined by OpenIoC and STIX formats. In addition, the authors also propose a technique knowledge graph (TKG), which is used to present higher-level techniques derived from various attack graphs. The approach is tested against 7373 procedures of 179 techniques obtained from MITRE ATT\&CK and 1515 CTI reports retrieved from diverse sources. While this approach does not use CVE information, it might be an interesting research topic to integrate ATT\&CK information into our approach. 

In \cite{li2022deepag} the authors present a completely different view on attack graphs. They try to find hints of possible attack steps in the system log and combine them into complete attack graphs. They developed the DeepAG framework that identifies threats and predicts possible attack paths based on system log entries. This is achieved by utilizing the transformer models, which allow for advanced persistent threat detection by modeling semantic information obtained from system logs. The approach applies Long Short-Term Memory (LSTM) that introduces bi-directional predictions. This led to the development of the mechanisms of Out-Of-Vocabulary (OOV) word processor allowing the incorporation of new attack patterns. The results show that over 99\% of 15000 sequences were detected by DeepAG. In contrast to our work, the goal of this work is not to identify possible vulnerabilities in advance, but to recognize a currently ongoing concrete attack. 

\cite{ZDEMIRSNMEZ2022102938} introduce Attack Dynamics, which is a novel tool that is able to automatically generate attack graphs based on CAPECs, CWEs, system topology, associated vendor products, and known vulnerabilities from metadata. Their tool is also able to link enterprise mitigations from the MITRE ATT\&CK framework to the security flaws it discovers. Furthermore, it is capable of integrating with an external optimization tool for security cost reduction both in real-time and offline, using a generated JSON output. In contrast, our approach is more bottom-up, in the sense that, we do not start on the most abstract level (CAPECs), but on the level of CVEs of the used components in a system. We then get more abstract by getting to the level of common weaknesses (CWEs). In Attack Dynamics, vulnerabilities must be added manually for each modeled component before they are considered in the further analysis and resulting modification of the attack graph. 

The approach of \cite{Khakpour2019} focuses on vulnerabilities of a self-adaptive system in transient states that occur especially during adaptations. Again, vulnerability information has to be added manually to the architecture description by utilizing Acme’s (\cite{Garlan00AcmeChapter}) properties of components. Finally, they use MulVAL (\cite{MulVAL2005}) to generate a probabilistic attack graph. The probabilities of the nodes are derived from the CVSS vectors of involved CVEs. For the definition of the parameters of the probability distribution for each leave of an attack tree, we utilize the epss framework and we do not focus on a specific type of system.

Unlike our work, \cite{Ou2006} generate attack trees, but focus on network-based models. The authors combine and analyze partial attack graphs created manually for different network nodes as Prolog rules in order to find possible successful attack paths. 

In summary, our bottom-up approach combines partially existing work in a new way and incorporates new developments, such as epss probabilities in order to generate attack trees using the proposed DSL grammar for all components of a given system configuration. The  vulnerabilities of a component are crawled automatically from public information security databases and are grouped by CWEs for each CPE using logical gates that are often applied fault trees. In addition, our approach automatically generates attack chains and combines the chains using \texttt{AND}, \texttt{SAND}, and \texttt{OR} gates, which are based on hierarchical relationships between weaknesses or CWEs. Moreover, the approach provides the possibility of generating more complex trees by considering CAPEC entries and semi-automatically relating attack chains using additional logic gates such as \texttt{PAND}, \texttt{SOR}, \texttt{FDEP}, \texttt{SPARE}, and \texttt{VOT} gates.

\section{Generation Approach}
\label{sec:approach}

In this section, the datasets that are used for the generation approach (Section \ref{sec:datasets}), the attack tree DSL language (Section \ref{sec:dsl}), and the generation approach itself (Section \ref{sec:generation}) are presented. The proposed approach involves designed artifacts, which include the DSL and fragment-based generation method. Thus, it takes into account the Design Science research principles (\cite{peffers2007design}). 

\subsection{Datasets}
\label{sec:datasets}

The National Vulnerability Database (NVD) CVE dataset, along with the CWE and CPE datasets from MITRE, provides a comprehensive resource for identifying and analyzing security vulnerabilities. The NVD CVE dataset contains over 150,000 entries, each detailing a unique vulnerability along with its severity rating, potential impact, and affected systems. The CWE dataset, on the other hand, contains more than 800 different types of software weaknesses. As for the CPE dataset, this includes more than 265,000 unique product names and versions.

These datasets are applied in a proposed approach for a comprehensive attack graph generation. It aims to identify all possible attack paths targeting a specific system. These were selected by considering the outcomes of the study on analysis and classification of public information security data sources by \cite{sauerwein2019analysis}. By leveraging the information provided in the CVE, CWE, and CPE datasets, it is possible to identify vulnerabilities and their root causes as well as the specific hardware and software products affected by each vulnerability (\cite{pekaric2021vulnerlizer}). In addition, the relationships between the datasets are utilized to generate connections between the specific vulnerabilities and calculate attack-related metrics. For obtaining CVE data, the NVD API is used, while CWE and CPE data are obtained from MITRE in the \texttt{.csv} data format.

\subsection{Domain Specific Language}
\label{sec:dsl}
In order to represent attack trees, we propose an attack tree DSL. 
The DSL is developed using the Xtext framework\footnote{\url{https://www.eclipse.org/Xtext/}}, which is the tool used for the development of programming languages and DSLs. 
The main reason behind choosing Xtext is that it allows a straightforward definition of DSLs. Furthermore, it is able to generate a fully functional editor for the Eclipse IDE, including features like auto-completion and syntax highlighting.  
All DSLs created with Xtext are based on EMF (\cite{Steinberg2008EMFEclipse}), which makes it possible to integrate the generated attack trees effortlessly with other models specified in an Xtext grammar. For example, attack trees can be integrated with fault trees as well as any other models using model transformations (\cite{witte22} and \cite{groner2023model}).
Listing~\ref{lst:att_tree_grammar} shows a simplified excerpt of the grammar that forms the base of our DSL for attack trees.

The root of an attack tree described in the proposed DSL is introduced by the keyword \texttt{AttackTarget}.
An \texttt{AttackTarget} provides several attributes to represent the different possible information provided by a CPE identifier for a specific library or platform.
Note that we defined special rules to represent a CPE, a CWE, and a CVSS vector to ensure that their specified structure is enforced, even for attack trees that are manually created using our DSL.   
Moreover, all attributes are optional and can occur in any sequence. This makes the creation of an attack tree more convenient and it is also possible to generate attack trees for which partial information is available. Listing~\ref{lst:att_tree1} presents a generated attack tree for the CPE \texttt{cpe:2.3:a:x.org:libx11:1.5.99.901:*:*\\:*:*:*:*:*}.
One can see that, e.g., the attribute description is not used in the specification of the \texttt{AttackTarget}.

The attribute \texttt{CVSS} of an \texttt{AttackTarget} is significant because it provides severity metrics related to CVEs.
This is particularly useful for combining attack trees with other models (\cite{kumar2017quantitative}). For example, the CIA values encoded in the CVSS vector can be used to decide whether an attack tree can be combined with a fault tree to derive an attack-fault tree.

The leaves of an attack tree are defined in our DSL by \texttt{AttackSteps}.
An \texttt{AttackStep} represents an individual vulnerability that is specified by a CVE entry.
Accordingly, an \texttt{AttackStep} offers the possibility to model the information provided by a CVE entry in our DSL through several optional attributes.
We have also defined the possibility of referencing attack steps in other ATs by their name, which is considered as an id. 
For example, the \texttt{AttackStep} in Listing~\ref{lst:att_tree1}, whose definition begins at line~11, can be referenced by other attack trees via its id \texttt{CVE202014344}.
This makes the definition of attack trees more convenient and modular.

Since a CPE identifier is linked to one or more CWEs, we offer the possibility to combine individual \texttt{AttackSteps} and subtrees using logical gates in our DSL.
As one can see in our grammar in Listing~\ref{lst:att_tree_grammar} in line~26, we offer \texttt{SAND}-gates in addition to the possibility to combine \texttt{AttackSteps} or subtrees by \texttt{AND}- and \texttt{OR}-gates.
\texttt{SAND}-gate defines that the incoming \texttt{AttackSteps} or subtrees must occur from left to right to activate the gate. 
In the attack tree in Listing~\ref{lst:att_tree1}, we use an \texttt{OR}-Gate to combine the two \texttt{AttackSteps}, which represent CWEs that are linked to the CPE \texttt{cpe:2.3:a:x.org:libx11:1.5.99.901:*:*\\:*:*:*:*:*}.

\begin{figure*} 
\begin{tiny}
\begin{lstlisting}[caption={Attack tree grammar},label={lst:att_tree_grammar}, language={AttackGrammar}, showstringspaces=false, numbers=left]
AttackTarget:
	"AttackTarget" (('id''='name=ID)? & ("CPE" "=" cpe=CPE)? & ("CWE" "=" cwe=CWE)? 
            &  ("CVSS" "=" cvss=CVSSVECTORList)? & ("note" "=" note=STRING)? & (("BaseScore" "=" baseScore=ScoreList)?) & 
            (("ImpactScore" "=" impactScore=ScoreList)?) & (("ExploitabilityScore" "=" exploitabilityScore=ScoreList)?)) 
            "{" attackTree=AttackTree "}"
;

CVSSVECTORList:
	'[' cvssList+=CVSSVECTOR (',' cvssList+=CVSSVECTOR)* ']'
;

ScoreList:
	'[' score+=REAL (',' score+=REAL)* ']'
;

AttackTree:
	step=AttackStep| subTree=SubTree| ref=[AttackTreeSubElements]
;

AttackStep: 
	"AttackStep" (name=ID)? (("description" "=" description=STRING) & (("CVE""="cve=CVE)?) & (("CVSS" "=" cvss=CVSSVECTOR)?)
            & (("probability""="probability=REAL)?) & (("BaseScore" "=" baseScore=REAL)?) & (("ImpactScore" "=" impactScore=REAL)?)
            & (("ExploitabilityScore" "=" exploitabilityScore=REAL)?) & (("epss" "=" epss=REAL)?) & (("note" "=" note=STRING)?))
;

SubTree:
	gate=Gate (name=ID)?  ("note" "=" note=STRING)?  "{" attackTree+=AttackTree (','attackTree+=AttackTree)* "}" 
;
Gate: 
	name='AND'| name='OR'| name='SAND' 
;
\end{lstlisting}
\end{tiny}
\end{figure*}

\subsection{Generation of Attack Trees}
\label{sec:generation}

The process of generating attack trees involves two primary steps. Initially, base trees are generated. Subsequently, more intricate trees are constructed by building upon the trees generated in the first step. In the remainder of this work, we will refer to these as base and derived attack trees. Python 3 was used to implement the generation tool, which has been made available in an open-source repository\footnote{\url{https://gitlab.com/uni4061424/attackgraphgen}}. 

\begin{figure*}[hbt!] 
\begin{minipage}{0.6\textwidth}
\begin{tiny}
\begin{lstlisting}[caption={A segment of the generated attack tree},label={lst:att_tree1}, language={AttackGraph}, showstringspaces=false, numbers=left]
AttackTarget
  CPE=cpe:2.3:a:x.org:libx11:1.5.99.901:*:*:*:*:*:*:* 
  CWE=CWE-190 
  cweNotes="Integer Overflow or Wraparound" 
  CVSS=[CVSS:3.1/AV:L/AC:L/PR:H/UI:N/S:U/C:H/I:H/A:H, 
        CVSS:3.1/AV:L/AC:L/PR:L/UI:N/S:U/C:H/I:H/A:H]
  BaseScore=[6.7, 7.8] 
  ImpactScore=[5.9, 5.9]
  ExploitabilityScore=[0.8, 1.8] 
  {
    OR {
      AttackStep 
        CVE202014344 
        description="An integer overflow leading to a heap-buffer
          overflow was found in The X Input Method (XIM) client [...]" 
        CVE=CVE-2020-14344 
        CVSS=CVSS:3.1/AV:L/AC:L/PR:H/UI:N/S:U/C:H/I:H/A:H 
        BaseScore=6.7 
        ImpactScore=5.9 
        ExploitabilityScore=0.8 
        epss=0.00051,
      AttackStep 
        CVE202014363 
        description="An integer overflow vulnerability leading to
          a double-free was found in libX11 [...]" 
        CVE=CVE-2020-14363 
        CVSS=CVSS:3.1/AV:L/AC:L/PR:L/UI:N/S:U/C:H/I:H/A:H 
        BaseScore=7.8 
        ImpactScore=5.9 
        ExploitabilityScore=1.8 
        epss=0.00042
      } 
  }

\end{lstlisting}
\end{tiny}
\end{minipage}
\begin{minipage}{0.39\textwidth}
\includegraphics[width=\textwidth]{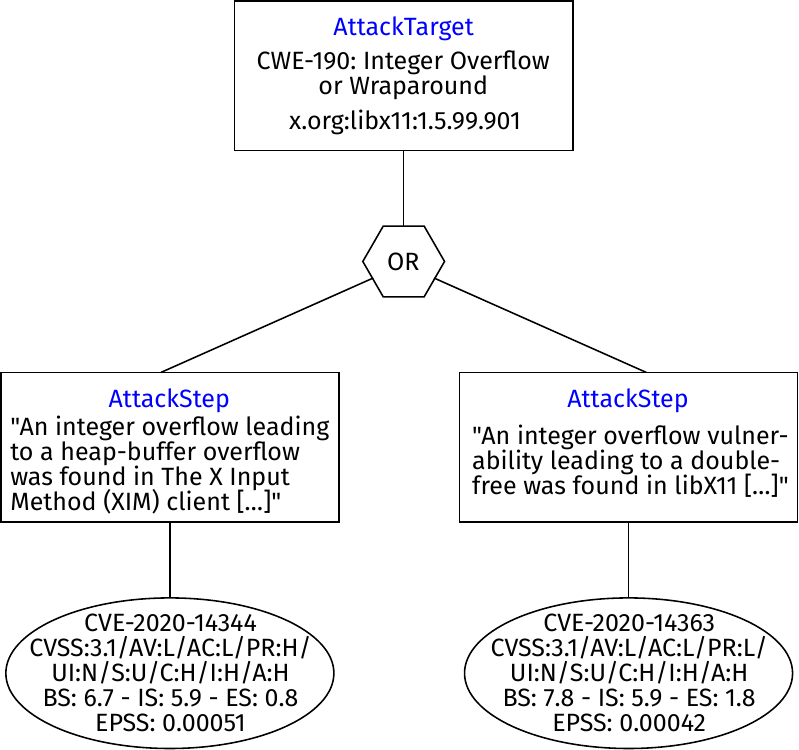}
\caption{Visualization of the generated attack tree segment}
\label{fig:attack_tree}
\end{minipage}
\end{figure*}

The base trees are generated strictly according to the relationship between the CVE, CPE, and CWE data. Since the input for the generation process is a list of CPE entries, it is necessary to devise which CVEs affect the addressed platforms. This is achieved through the many-to-many relationship between CPEs and CWEs as well as the many-to-many relationship between CWEs and CVEs. As a result, an individual file is generated for each CPE in which there are as many trees as there are CWEs for that particular CPE. In other words, the number of trees per CPE equals the number of related CWEs and these are represented with the \texttt{AttackTarget} node. The reason behind this is to obtain "simple" base trees for each specific CPE, allowing them to be used as building blocks for derived trees. Consequently, the base trees are generated according to the proposed DSL grammar described in Section \ref{sec:dsl}. For example, for the CPE \texttt{cpe:2.3:a:x.org:libx11:1.5.99.901:*:*\\:*:*:*:*:*} (libx11 library, version 1.5.99.901) and \texttt{CWE-190} (Integer Overflow or Wraparound), the attack tree presented in Listing \ref{lst:att_tree1} is generated. An equivalent graph-like visualization of the attack tree segment is shown in Figure~\ref{fig:attack_tree}.

This attack tree represents just one of the generated trees. The other trees as well as some additional notes are omitted for presentation purposes. The tree includes two vulnerabilities namely \texttt{CVE-2020-14344} and \texttt{CVE-2020-14363}, which include CVSS and epss scores (Exploit Prediction Scoring System) scores\footnote{\url{https://www.first.org/epss/}}. These are connected using the \texttt{OR} gate and each CVE is represented as a leaf of the tree (\texttt{AttackStep} node). The CVEs are directly related to the presented CPE via the \texttt{CWE-190}. This means that an adversary can exploit any of the two vulnerabilities in order to attack the target system. During the generation of base trees, the only gate that is used is the \texttt{OR} gate. This is because each vulnerability is a standalone and can be part of a single attack. However, in the second generation step, these are used as the main building blocks of attack chains that are part of derived trees. Regarding the scores and metrics of CVEs within the tree, these are propagated to the root of the tree (\texttt{AttackTarget}). The propagation is conducted for each score, which includes the \texttt{BaseScore}, \texttt{ImpactScore}, and \texttt{ExploitabilityScore} as well as the CVSS vector. This is done for both CVEs due to the possibility of using such metrics when integrating attack trees with other models such as fault trees. Since the \texttt{OR} gate is used, only the scores related to a single CVE will be used. However, this can only be known once the model integration is executed. The reason why the epss scores are not propagated is that these scores relate to probability over time and it would not make sense to propagate them towards the top of the tree.

Similarly to base trees, derived trees are also generated for a specific CPE. The goal of derived trees is to generate possible attack chains for a specific software or software library. In order to generate these links, the relationship information from the CWE database was used. This includes \textit{ChildOf}, \textit{ParentOf}, \textit{RequiredBy}, \textit{Requires}, \textit{CanFollow}, and \textit{CanPrecede} relationships\footnote{\url{https://cwe.mitre.org/data/xsd/cwe_schema_v6.10.xsd}}. The \textit{ChildOf} and \textit{ParentOf} relationships in the CWE are used to show all the hierarchical relationships. The \textit{Requires} relationship indicates that a certain weakness or vulnerability is needed for another weakness to exist or be exploited. Similarly, the \textit{RequiredBy} relationship denotes that a certain weakness or vulnerability is a prerequisite for another weakness to be present or exploitable. The \textit{CanFollow} relationship indicates that one weakness can occur after another in a sequence of events, while the \textit{CanPrecede} denotes that one weakness can occur before another in a sequence of events.  

According to the aforementioned relationships including the CWE-CVE relationship and the DSL grammar, the attack tree generator forms additional links between all the CVEs that were generated as base trees for a particular CPE. In addition to the existing \texttt{OR} gate relations, CVEs can now be related by \texttt{AND} and \texttt{SAND} gates based on the \textit{RequiredBy}, \textit{Requires}, \textit{CanFollow}, and \textit{CanPrecede} relationships respectively. Namely, if a CWE \texttt{a} is related to a CWE \texttt{b} via the \textit{RequiredBy} relationship, each CVE from the CWE \texttt{a} is also related to CVEs from the CWE \texttt{b}. As a result, these are connected using the \texttt{AND} gate as a part of the newly generated tree. In regards to CVSS scores propagation, this is done differently for \texttt{SAND} and \texttt{AND} gates compared to the \texttt{OR} gate. This is due to the fact that each \texttt{AttackStep} or CVE needs to be exploited to activate the gate. As a result, the mean values for the CVSS scores (\texttt{BaseScore}, \texttt{ImpactScore}, and \texttt{ExploitabilityScore}) are propagated to the \texttt{AttackTarget} node. However, all CVSS vectors are propagated (as in the \texttt{OR} gate case) because it is not possible to calculate the mean values of some metrics within the vector, such as the Attack Vector (AV) metric. This metric provides the context in which vulnerability exploitation is possible.

The generation of attack chains is conducted recursively by continuously checking relations between vulnerabilities and forming new chains. In order to prevent extensive generation, especially in the case of CPEs with a large number of related CWEs and CVEs, it is possible to specify the largest depth of the generated attack trees. The implementation also restricts loops in attack chains by making sure a specific CVE is not part of a single chain more than once. Listing \ref{lst:att_tree2} presents an example of the generated derived tree wherein the \texttt{SAND} gate was used to demonstrate an attack chain. The attack chain includes \texttt{CVE-2020-14344} and \texttt{CVE-2021-31535} for which the mean values of the CVSS scores are calculated. The following section presents a demonstrator example showing how an existing attack chain can be generated using the proposed approach.

\begin{tiny}
\begin{lstlisting}[caption={Generated derived tree},label={lst:att_tree2}, language={AttackGraph}, showstringspaces=false, numbers=left]https://www.overleaf.com/project/6420612a43950034224578ad
AttackTarget 
    CVSS=[CVSS:3.1/AV:L/AC:L/PR:H/UI:N/S:U/C:H/I:H/A:H,
          CVSS:3.1/AV:N/AC:L/PR:N/UI:N/S:U/C:H/I:H/A:H]
    BaseScore=[8.3] ImpactScore=[5.9] ExploitabilityScore=[2.4] {
    SAND S1 {
        CVE202014344,
        CVE202131535
    } 
}
\end{lstlisting}
\end{tiny}

\section{Discussion}
\label{sec:disseval}
In this section, we provide a demonstrator example (Section \ref{sec:case}), implications (Section \ref{sec:discussion}), and limitations of the proposed approach (Section~\ref{sec:threats}).
\subsection{Working Example}
\label{sec:case}

In order to demonstrate the effectiveness of the proposed attack tree generator, we attempt to recreate an existing attack chain that was identified by the cybersecurity community. The description of the CVE entry \texttt{CVE-2019-9500} states the following: \textit{[...]. This vulnerability can be exploited with compromised chipsets to attack the host, or when used in combination with CVE-2019-9503, can be used remotely [...]} By conducting a simple check on the NVD page, it can be seen that both of these vulnerabilities are associated with the \texttt{cpe:2.3:a:broadcom:brcmfmac\_driver:-:*\\:*:*:*:*:*:*} CPE, which relates to the Broadcom brcmfmac driver.

%
\begin{figure}[hbt!]
\begin{tiny}
\begin{lstlisting}[caption={Demonstrator example},label={lst:att_tree3}, language={AttackGraph}, showstringspaces=false, numbers=left]
AttackTarget 
  CVSS=[CVSS:3.1/AV:A/AC:H/PR:N/UI:N/S:C/C:H/I:H/A:H,
        CVSS:3.1/AV:A/AC:H/PR:N/UI:N/S:C/C:H/I:H/A:H]
  BaseScore=[8.3] ImpactScore=[6.0] ExploitabilityScore=[1.6] {
  SAND S0 {
      AttackStep 
        CVE20199503 
        description="The Broadcom brcmfmac WiFi driver ..." 
        CVE=CVE-2019-9503 
        CVSS=CVSS:3.1/AV:A/AC:H/PR:N/UI:N/S:C/C:H/I:H/A:H 
        BaseScore=8.3 
        ImpactScore=6.0 
        ExploitabilityScore=1.6
        epss=0.00155
    },
      AttackStep
        CVE20199500 
        description="The Broadcom brcmfmac WiFi driver ..." 
        CVE=CVE-2019-9500 
        CVSS=CVSS:3.1/AV:A/AC:H/PR:N/UI:N/S:C/C:H/I:H/A:H 
        BaseScore=8.3 
        ImpactScore=6.0 
        ExploitabilityScore=1.6
        epss=0.00683
    }
  } 
}
\end{lstlisting}
\end{tiny}

\includegraphics[width=\linewidth]{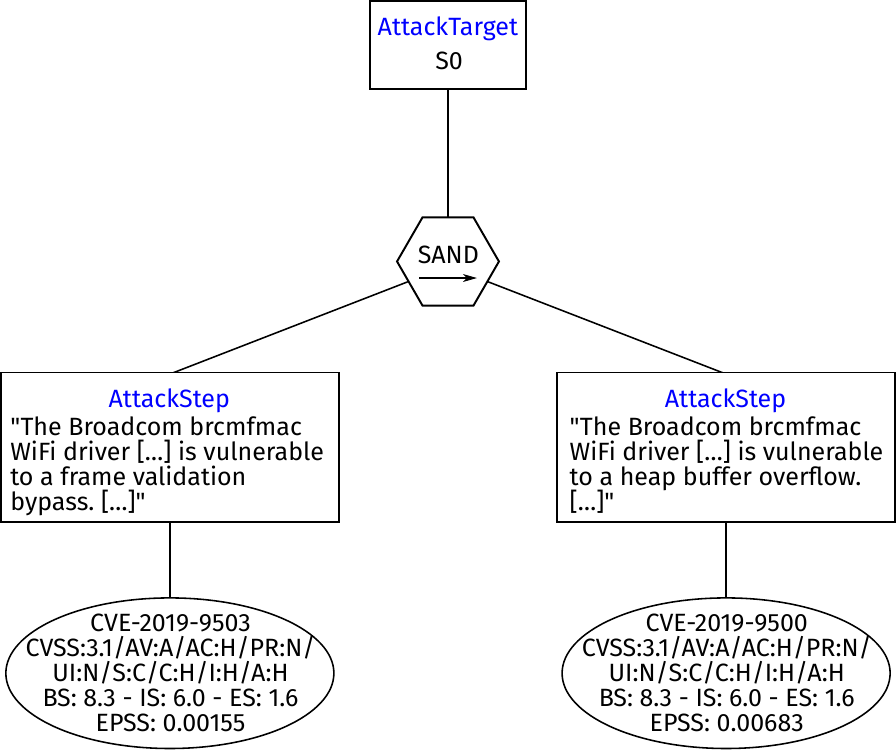}
\caption{Visualization of the demonstrator example}
\label{fig:example}
\end{figure}

Listing \ref{lst:att_tree3} presents the generated output for the aforementioned CPE, while Figure \ref{fig:example} provides its visualization. The generated tree includes the two CVEs that are represented as separate \texttt{AttackSteps}, which belong to two different CWEs (\texttt{CWE-20} and \texttt{CWE-122}). The relation between \texttt{CVE-2019-9500} and \texttt{CWE-787} was omitted from the tree for presentation purposes. For the same reason, the CVE tags are not used as in Listing \ref{lst:att_tree2}. The third \texttt{AttackTarget} represents the attack chain that was to be recreated. It connects both CVEs using the \texttt{SAND} gate, which means that they can be either exploited in a sequence from left to right or completely standalone. In other words, improper input validation may lead to a heap-based buffer overflow. This specific use case was chosen as it does not involve too many CVEs and attack chains, making its presentation more appealing. 

\subsection{Implications}
\label{sec:discussion}

In the following, we discuss the possible implications of the presented attack tree generation approach. Since the total number of CVEs is constantly increasing, this leads to the generation of complex attack graphs. As such, it requires significant computation power and time to generate. However, there are benefits to generating attack graphs from fragments, as it allows for a focus on specific weaknesses, libraries, and high-severity vulnerabilities. These fragments can later be connected to form more complex attacks if necessary. This can be done semi-automatically by manually developing attack patterns and considering existing patterns from the CAPEC database. According to this, it is also possible to consider other types of gates such as \texttt{PAND}, \texttt{SOR}, \texttt{FDEP}, \texttt{SPARE}, and \texttt{VOT} gates when combining generated fragments.  

Despite the potential benefits, the fully automated generation of attack graphs from fragments is challenging due to the missing fields and relationships within public security databases. However, this approach allows for a more targeted and efficient analysis of vulnerabilities. By focusing on specific weaknesses and software libraries, analysts can prioritize the most critical vulnerabilities and develop more effective mitigation strategies. This approach can also help identify dependencies between different vulnerabilities. This can be used to understand potential attack paths better and develop more effective defense mechanisms. One possible challenge with this approach is the potential for false positives or false negatives. When generating individual tree fragments, it is possible to miss meaningful connections or relationships between vulnerabilities. However, combining fragments in a semi-automated manner makes it possible to mitigate this risk and develop more comprehensive attack graphs. 


Moreover, the semi-automated approach to generating attack graphs from fragments has shown promise in mitigating vulnerabilities in complex systems such as self-adaptive and AI systems. These types of systems are characterized by continuous adaptation and change, including changes in the software and libraries they use. In such cases, the re-generation of attack trees can become highly complex. However, the use of fragments in attack graph generation allows for the regeneration of only the affected fragments. This significantly reduces the time and computing power required. In addition, it is possible to generate trees for specific components of a system and later, if necessary, integrate these standalone trees. However, attacks that target AI logic would have to be manually integrated into the generated attack trees. Overall, this approach provides a solid foundation for future research in this direction, enabling more comprehensive threat analysis and vulnerability management in complex systems. While the semi-automated approach may require significant expertise, its potential benefits make it a valuable tool in advanced threat analysis and vulnerability management for any security practitioner including security experts from both industry and academia.

\subsection{Limitations}
\label{sec:threats}

One of the significant limitations is the incomplete data on publicly available information security databases. In the proposed approach, only the currently available data was utilized. As there are no guarantees of completeness of such data, this can lead to incomplete or biased research outcomes. Additionally, there are threats in terms of the consistency and accuracy of the data as different organizations maintain these databases. As a result, there is no standardization in terms of data entry and management. This makes it difficult for researchers to perform accurate and comprehensive analyses using such data. Therefore, there is a need for a standardized approach to maintaining and managing these databases. This can help ensure the completeness, consistency, and accuracy of the data, which would potentially lead to more reliable research outcomes. However, this requires significant effort from the whole cybersecurity community.

Another limitation in the software supply chain is matching CPE data to software and libraries. This is due to various reasons, such as a growing number of software and libraries including the large number of versions as well as inconsistent naming of CPE data. Regarding the generation of attack chains, multiple limitations exist that mostly relate to missing dependencies between CWE data. This is due to the fact that many entries do not contain any relationships, which could potentially result in the failed detection of relevant attack chains. The proposed approach for attack graph generation can be considered generalizable as it is not limited to specific types of systems regardless of their type or complexity. 



\section{Conclusion}
\label{sec:conclusion}
There is a growing complexity in CPSs, including those that necessitate self-management capabilities as well as those enabled by machine learning and AI. This nuance requires advanced security assurance techniques that are efficient and scalable. In this paper, we have presented a fragment-based approach for security threat modeling with attack graphs for complex cyber-physical systems, which can be utilized for other types of systems too. The approach relies on defining an attack tree model, using a proposed DSL grammar specified within the Xtext framework to generate simple attack trees. Consequently, they can be combined based on defined attack patterns into more sophisticated attack chain models.

Future work involves further improving the attack tree generator. This includes conducting NLP analysis on security data, which would allow the identification of additional attack chains and the use of additional gates to create more complex attack trees. For instance, this can be done by analyzing CVE descriptions in more detail as well as using CAPEC and manually crafted attack patterns. By doing so, this can guide the formation of attack chains that include exploits on distinct software and libraries. Furthermore, additional empirical studies will be conducted to further enhance the efficiency and effectiveness of the generated trees. Further evaluation is planned that will involve the design of model-checkers that verify generated attack trees. In addition, the proposed approach will be tested in real-case scenarios together with our partners from the industry.

\section*{Acknowledgements} 
This work was partially supported by the Austrian Science Fund (FWF): I 4701-N and the German Research Foundation (DFG): 435878599. We also thank the Hilti Corporation for their support.

\printbibliography[title={\section{References} 
\label{sec:references}}]
\end{document}

%% file: hicss-packages.tex
\usepackage[letterpaper]{geometry}
\usepackage{hicss}
\usepackage{times}
\usepackage[none]{hyphenat}
\usepackage{url}
\usepackage{latexsym}
\usepackage{indentfirst}
\usepackage{graphicx}
\graphicspath{{images/}}
\usepackage[style=apa]{biblatex}
\AtNextBibliography{\small}